\documentclass[runningheads,a4paper]{llncs}[2015/06/24]

\usepackage{cmap}
\usepackage[T1]{fontenc}

\usepackage{graphicx}

\usepackage{pdfpages}

\usepackage[ngerman,english]{babel}
\addto\extrasenglish{\languageshorthands{ngerman}\useshorthands{"}}

\usepackage[%
rm={oldstyle=false,proportional=true},%
sf={oldstyle=false,proportional=true},%
tt={oldstyle=false,proportional=true,variable=true},%
qt=false%
]{cfr-lm}
\usepackage[math]{blindtext}

\usepackage{amsmath}

\usepackage{cite}

\usepackage{paralist}

\usepackage{csquotes}

\usepackage{microtype}

\usepackage{url}
\makeatletter
\g@addto@macro{\UrlBreaks}{\UrlOrds}
\makeatother

\usepackage{booktabs} 

\usepackage{color, colortbl}
\usepackage{xcolor}
\definecolor{darkblue}{rgb}{0,0,.5}
\definecolor{Brown}{cmyk}{0,0.81,1,0.60}
\definecolor{OliveGreen}{cmyk}{0.64,0,0.95,0.40}
\definecolor{CadetBlue}{cmyk}{0.62,0.57,0.23,0}
\definecolor{darkgray}{rgb}{0.5, 0.5, 0.5}
\definecolor{lightlightgray}{gray}{0.9}
\definecolor{lightgreen}{rgb}{0.67, 1, 0.32}
\definecolor{EsiOrange}{HTML}{F39600}
\definecolor{EsiBlue}{HTML}{006d9a}
\definecolor{EsiBlue2}{HTML}{006d9a}
\definecolor{EsiGreen}{HTML}{9ac013}
\definecolor{AimesGreen}{HTML}{42A83B}
\definecolor{AimesBlue}{HTML}{31A5F3}
\definecolor{bubblegum}{rgb}{0.99, 0.76, 0.8}
\definecolor{palegreen}{rgb}{0.6, 0.98, 0.6}
\definecolor{blond}{rgb}{0.98, 0.94, 0.75}

\usepackage{pdfcomment}
\hypersetup{hidelinks,
   colorlinks=true,
   allcolors=black,
   pdfstartview=Fit,
   breaklinks=true}
\usepackage[all]{hypcap}

\usepackage[capitalise,nameinlink]{cleveref}
\crefname{section}{Sect.}{Sect.}
\Crefname{section}{Section}{Sections}

\usepackage{xspace}

\DeclareFontFamily{U}{MnSymbolC}{}
\DeclareSymbolFont{MnSyC}{U}{MnSymbolC}{m}{n}
\DeclareFontShape{U}{MnSymbolC}{m}{n}{
    <-6>  MnSymbolC5
   <6-7>  MnSymbolC6
   <7-8>  MnSymbolC7
   <8-9>  MnSymbolC8
   <9-10> MnSymbolC9
  <10-12> MnSymbolC10
  <12->   MnSymbolC12%
}{}
\DeclareMathSymbol{\powerset}{\mathord}{MnSyC}{180}

\usepackage{listings}
\begin{document}

\pdfinclusioncopyfonts=1

\title{Hybrid Fortran: High Productivity GPU Porting Framework Applied to Japanese Weather Prediction Model}

\titlerunning{Hybrid Fortran: High Productivity GPU Porting}

\author{Michel M\"{u}ller \and Takayuki Aoki}
\institute{Tokyo Institute of Technology}



\maketitle

\begin{abstract}
In this work we use the GPU porting task for the operative Japanese weather prediction model ``ASUCA'' as an opportunity to examine productivity issues with OpenACC when applied to structured grid problems. We then propose ``Hybrid Fortran'', an approach that combines the advantages of directive based methods (no rewrite of existing code necessary) with that of stencil DSLs (memory layout is abstracted). This gives the ability to define multiple parallelizations with different granularities in the same code. Without compromising on performance, this approach enables a major reduction in the code changes required to achieve a hybrid GPU/CPU parallelization - as demonstrated with our ASUCA implementation using Hybrid Fortran.
\end{abstract}


\section{Introduction} \label{sec:intro}

With supercomputing shifting towards accelerators, the increasing need for manycore architecture support in software has created a divide between domain scientists who mainly care about modeling, and the supercomputers their applications are required to run on. The high rate of change in both hardware and software creates maintainability issues, especially with codes that are required to run on varying hardware architectures such as multi-core CPU and GPU. The aforementioned divide has widened especially in atmospheric sciences, a field with constant need for model adaptations and high demand for computing resources.
Related publications suggest that productivity issues are what is holding back wider accelerator support in this field.

\medskip

In this work we use the GPU porting task for a Japanese weather prediction model (``ASUCA'') as an opportunity to examine productivity issues with the current GPGPU standard ``OpenACC'' when it is applied to structured grid problems written in Fortran. ASUCA is the main mesoscale weather prediction model developed at the Japan Meteorological Agency. It is used in operation, generating nine-hour-forecasts every hour \cite{sakamotodevelopment}\cite{ishida2010development}.

\medskip

We then propose Hybrid Fortran, a solution that is designed to increase productivity when re-targeting structured grid Fortran applications to GPU. It is an improvement over OpenACC in two major aspects: 

\begin{enumerate}
    \item Parallelization granularity is abstracted. This allows the user to have multiple granularities defined in the same codebase, depending on the targeted hardware architecture. This is a crucial advantage in order to implement ASUCA's physical processes on GPU - a code that originally has a very coarse granularity, which is ill-matched for GPUs.
    \item Memory layout is abstracted while supporting the already existing user code. More specifically, the layout is reordered at compile-time to match the target architecture, and extended with additional dimensions to match the specified parallelization granularity.
\end{enumerate}

\medskip

By investigating the necessary code changes with a completed implementation based on Hybrid Fortran we show that this method has enabled high productivity and performance for re-targeting ASUCA to GPU. More than 85\% of the hybridized codebase is a one-to-one copy of the original CPU-only code - without counting white-space, code comments and line continuations. An equivalent OpenACC-based solution of ASUCA is estimated to require more than ten thousand additional code lines, or 12.5\% of the reference codebase. The new implementation performs up to 4.9x faster when comparing one GPU to one multi-core CPU socket. On a full-scale production run with 1581 x 1301 x 58 grid size and 2km resolution, 24 Tesla P100 GPUs are shown to replace more than 50 18-core Broadwell Xeon sockets.

\medskip

This paper is structured as follows: Section \ref{sec:asuca} introduces the application our work focuses on. In Sections \ref{sec:granularity} and \ref{sec:layout} we introduce the main difficulties that we face when porting this application to GPU. Section \ref{sec:existing} outlines the related work in terms of existing methods to solve these difficulties. We then provide a problem summary in Section \ref{sec:problem}. In Sections \ref{sec:hybridFortran}, \ref{sec:hf-transformation} and \ref{sec:results} we discuss our solution to this problem, the underlying code transformation method, as well as productivity and performance results achieved with this solution, respectively. Finally, in Section \ref{sec:conclusions} we draw conclusions and point out future work. 

\subsection{ASUCA on GPU} \label{sec:asuca}

The regional scale weather prediction model ``ASUCA'' is one of the main operational forecast models in Japan \cite{ishida2010development}. It is developed by the Japan Meteorological Agency (JMA) and used in production since 2014, covering all of Japan as well as relevant ocean areas and surrounding landmasses in East Asia on a rectangular grid with two kilometer resolution, using the finite volume method for the spatial discretization \cite{shimokawabe:2014}.

\medskip

ASUCA is implemented in Fortran, with multidimensional arrays stored in modules as its main data structure. It is structured as a dynamical core interfacing with physical processes, as commonly seen in other weather models such as WRF \cite{wrf_on_phi} and COSMO \cite{cummings:review}.

\medskip

Parallelization is applied in the horizontal domain (iterated with \verb|I| and \verb|J| loop indices). In order to scale to multiple nodes, ASUCA's grid is decomposed into blocks in the horizontal domain that are scattered across nodes, thus requiring halo communication. OpenMP parallel loop directives are employed as an existing intra-node CPU parallelization. Time discretization is implemented by employing a third order Runge-Kutta scheme, enabling long time steps \cite{wicker2002time}. Sound and gravity waves are treated separately using a second-order Runge-Kutta scheme to enable a higher time resolution, employing the HEVI scheme (Horizontally explicit - vertically implicit) \cite{shimokawabe:2014}.

\medskip

ASUCA's dynamical core is a stencil code and thus heavily bounded by memory bandwidth \cite{shimokawabe201080} \cite{dursun2009core}. Since the dynamical core also dominates the runtime, GPUs are an attractive target architecture, with a memory bandwidth that is typically 5 to 7 times higher than Intel Xeon architectures of a similar generation. This work is thus motivated by the task of achieving a GPU port for ASUCA, however with the additional goal of optimizing for minimal code changes in order to achieve better acceptance from the application owner (JMA).

\medskip

ASUCA is a protected asset of the Japanese government and can not be published at this time - in this paper we thus refer to code snippets to discuss our implementation.

\subsection{Parallelization Granularity} \label{sec:granularity}

Compared to CPUs, GPUs support a very high number of parallel threads while having a very low thread switching overhead - however with the cost of small caches available per thread and a low single-threaded performance. Furthermore, the latency experienced when off-loading code to GPU results in a very high number of threads being optimal, preferably a multiple of the available arithmetic units (i.e. tens of thousands of threads or more). This in turn leads to register pressure being a major factor for the optimization of GPU code, since the number of utilized registers scales linearly with the number of active threads. These characteristics point to an important distinction in the GPGPU programming model: A fine-grained parallelization is strongly preferred or even necessary (as subroutine calls in kernels are required to be inlined, which becomes impractical with deep call graphs).

\medskip

While ASUCA's original implementation already offers a fine-grained and thus GPU-friendly dynamical core, most of its physical processes use very large kernels and thus coarse granularity. Since each vertical column (\verb|K| index) can be computed independently for many of the physical processes, these computations are implemented in a single kernel in order to increase cache locality and decrease the amount of context switching and thread synchronization \cite{kwiatkowski2001evaluation}\cite{douglas2000cache}.

\medskip

In order to achieve GPU support, a more fine-grained parallelization is thus required for the physical processes. An automated or assisted approach for kernel fission is desirable in order to allow for GPU acceleration while keeping cache locality and low overhead for the CPU case.

\subsection{Memory Layout} \label{sec:layout}

There are two major aspects in which the memory layout on GPU differs from that on CPU: 

\subsubsection{Stride-1 Access} On CPU, the memory layout is generally chosen such that the fastest varying dimension is mapped to stride-1 access. GPUs on the other hand require the first parallel dimension to be mapped to stride-1 in order to coalesce the memory operations, since a group of threads is executed in lockstep. In case of ASUCA these are not the same dimensions: \verb|K| is the fastest varying dimension, yet it is executed sequentially in general (with a few possible exceptions in the dynamical core), while either \verb|I| or \verb|J| can be chosen as the first parallel dimension. The original codebase thus employs \verb|KIJ| memory order. In an unpublished paper submitted for review, we show that \verb|KIJ| order leads to a 7.7x slowdown on GPU (versus \verb|IJK|) while \verb|IJK| ordering leads to a 35\% slowdown on CPU (versus \verb|KIJ|) \cite{mueller2017unpub}.

\subsubsection{Privatization} To create a more fine-grained parallelization for GPU, as discussed in Section \ref{sec:granularity}, kernel fission is required. This in turn requires thread-local data structures and passed-in data slices to be extended in the parallel domain (\verb|IJ|). Many scalars thus become 2D-arrays. Many 1D- become 3D-arrays.

\medskip

Thus, in order to make performance portability possible, the memory layout is required to be changeable and extendable, depending on the target architecture and the parallelization granularity. To achieve this in pure Fortran however, all array specifications and accesses need to be modified and thus specialized for GPU (i.e. code duplication is necessary).

\subsection{Related Work} \label{sec:existing}

The following works are related to this paper in that they describe alternative or similar methods to achieve a hybrid GPU/CPU capable port for atmospheric models. All of these works have achieved compelling speedups on GPU, this discussion therefore focuses on the productivity aspects. 

\subsubsection{Stencil Domain-Specific Languages}

Domain-specific languages applied to stencil algorithms have been one method to abstract parallelization boiler-plate and memory layout for hybrid GPU/CPU code. For this matter, direct data accesses with loop- or thread indices are abstracted in the point-wise stencil code. This generally requires a complete rewrite of existing code. We take note of the following implementations of weather models using this technique:

\begin{itemize}
\item Shimokawabe et al. have completed a research implementation of the ASUCA dynamical core and a portion of its physical processes using their own C++ stencil DSL library \cite{shimokawabe:2014},
\item Fuhrer et al. have implemented the dynamical core of COSMO for operational use at MeteoSwiss, using a purpose-built C++ stencil DSL library for structured grid applications (``STELLA'') \cite{fuhrer2014towards}. The successor project currently enhancing this method is called ``GridTools'' \cite{GridTools}.
\item Jumah et al. have proposed a general grid definition and manipulation language (GGDML), an extension to Fortran with applicability to other languages, based on the requirements for three existing models: DYNAMICO, ICON and NICAM. This work is part of the AIMES project \cite{jumah2017ggdml}.
\end{itemize}

\subsubsection{Directive-Based Porting Methods}

Directives are used to steer compilers on how to optimize or parallelize already existing code for a specific hardware architecture. Privatization of thread-local data is generally supported in directive-based methods, while storage ordering is not. To switch between multiple parallelization granularities, code duplication is necessary. We are aware of the following implementations of atmospheric models with this method: 

\begin{itemize}
\item Lapillonne et al. have implemented the relevant physical processes of COSMO for operational use at MeteoSwiss using OpenACC in Fortran \cite{lapillonne2014using}.
\item Govett et al. have ported the dynamical core of the Fortran-based Non-hydrostatic Icosahedral Model (NIM) to GPU, first using their own directive-based transformation tool ``F2C-ACC'' and later using OpenACC (after critical performance issues and bugs were fixed by the compiler vendors). At the time of this writing we are not aware of a GPU implementation of the NIM physical processes however, an issue that has reduced the potential speedup due to the communication overhead caused by running the physical processes on CPU \cite{govett2014directive}\cite{govett2017parallelization}.
\item Norman et al. have implemented the ``Accelerated Model for Climate and Energy'' (ACME) for the U.S. DOE using OpenACC. As with ASUCA, ACME's physical processes are problematic for GPU due to their coarse-grained parallelization. GPU-specific code duplication was the only solution found when using OpenACC \cite{norman2017exascale}.
\end{itemize}

\subsubsection{Granularity Optimization Methods}

Kernel fusion has been the main approach to granularity optimization applied to GPGPU programming we are aware of. We take note of the following related work:

\begin{itemize}
\item Wahib and Maruyama have shown the effectiveness of this approach in terms of performance when applied to CUDA C kernels \cite{wahib}.
\item Gysi and Hoefler have applied the same approach as well as loop fusion for the aforementioned STELLA stencil DSL library \cite{gysiintegrating}.
\item Clement et al. are applying kernel fusion and an OpenACC/OpenMP targeted transformation to Fortran code in an ongoing effort as part of the C2SM project (as of yet unpublished\footnote{These open-sourced efforts can be found at \url{https://github.com/C2SM-RCM/claw-compiler}.}).
\end{itemize}

We are not aware of previous or ongoing work regarding kernel fission (and thus support for coarse-grained parallel programming) applied to GPUs.

\subsubsection{Memory Layout Abstraction Methods}

While stencil DSLs abstract the memory layout, they also require a full rewrite of the point-wise code. The following work allows for the reuse of existing code while keeping the performance portability gains of an abstracted memory layout:

\begin{itemize}
    \item With the C++ library ``Kokkos'' (part of the Trilinos project), Edwards et al. have demonstrated that existing point-wise code can be reused even when the underlying data structures are converted to an abstracted memory layout. OpenMP, Pthreads as well as CUDA are provided as backends to the user code for this library. Granularity optimizations are not supported at the time of this writing, neither is Fortran user code \cite{CarterEdwards20143202}.
    \item With a DSL created for the climate model ``ICON'', Torres et al. have shown that the Fortran syntax can be extended to allow for an abstracted memory layout. A code transformation based on the ROSE compiler\cite{quinlan2000rose} is employed towards that goal \cite{torres2013icon}. Sawyer et al. have subsequently built on this abstraction to port the dynamical core of ICON to GPU using OpenACC directives \cite{sawyer2014towards}. We are not aware of any granularity optimizations supported in the ICON DSL.
\end{itemize}

\subsection{Problem Summary} \label{sec:problem}

No existing method, that we are aware of, combines memory layout abstraction and a flexible parallelization granularity with the ability to reuse existing Fortran code for GPGPU. In this work we aim at introducing such a method.








\section{Hybrid Fortran Language Extension and Code Transformation} \label{sec:hybridFortran}

Hybrid Fortran has been developed as a method for porting structured grid Fortran applications to GPU. In recognition of the advantages and disadvantages of stencil DSL- and directive-based methods (outlined in Section \ref{sec:problem}) we have combined the advantages of both by employing the following characteristics in our approach: 

\begin{enumerate}
 \item \label{enum:parallelization} it \textit{does abstract} the parallel loops in order to achieve multiple parallelization granularities with the same code,
 \item \label{enum:point-wise} it \textit{does not abstract} the point-wise code (i.e. the loop bodies) - allowing for code reuse,
 \item \label{enum:layout} it \textit{does separate} the memory layout as defined in the user code from the layout that is effectively implemented for each architecture.
\end{enumerate}

Hybrid Fortran is an open-source framework and can be accessed together with a library of sample applications\footnote{Please refer to \url{https://github.com/muellermichel/Hybrid-Fortran}.}.

\medskip

In this Section we discuss how these characteristics have been achieved. Subsection \ref{sec:hf-parallelization} describes our approach to parallelization and granularity in Hybrid Fortran, while Subsection \ref{sec:hf-layout} discusses and compile-time defined memory layout and device memory handling. 

\subsection{Parallel Loop Abstraction} \label{sec:hf-parallelization}

Consider the following kernel from JMA's ASUCA reference implementation. As part of the dynamical core it is executed within the second-order Runge-Kutta scheme with high time resolution. It applies lateral and upper damping to ASUCA's grid point values.

\begin{lstlisting}[name=lateral_damping, label=listing:lateral_damping, caption={Lateral and upper damping kernel applied to grid point values.}]
!$OMP PARALLEL DO
  do j = ny_mn, ny_mx
  do i = nx_mn, nx_mx
  do k = nz_mn, nz_mx
    dens_ptb_damp(k,i,j) = &
      &  mtratio_bnd * ( dens_ref_f(k,i,j) + dens_ptb_bnd(k,i,j,1) ) &
      & + tratio_bnd * ( dens_ref_f(k,i,j) + dens_ptb_bnd(k,i,j,2) ) &
      & - dens_ref_f(k,i,j)
  end do
  end do
  end do
!$OMP END PARALLEL DO
\end{lstlisting}

Using Hybrid Fortran we replace the OpenMP directives, as well as the loop instructions to be parallelized,  with our parallelization DSL: 

\begin{lstlisting}[name=lateral_damping_hf, label=listing:lateral_damping_hf, caption={Lateral and upper damping kernel, modified with Hybrid Fortran.}]
@parallelRegion{ &
  & domName(i,j), domSize(nx_mn:nx_mx,ny_mn:ny_mx), &
  & startAt(nx_mn,ny_mn), endAt(nx_mx,ny_mx), template(TIGHT_STENCIL) &
& }
do k = nz_mn, nz_mx
    dens_ptb_damp(k,i,j) = &
      &  mtratio_bnd * ( dens_ref_f(k,i,j) + dens_ptb_bnd(k,i,j,1) ) &
      & + tratio_bnd * ( dens_ref_f(k,i,j) + dens_ptb_bnd(k,i,j,2) ) &
      & - dens_ref_f(k,i,j)
end do
@end parallelRegion
\end{lstlisting}

We therefore have an explicit distinction between loops that are treated as parallelizeable (and are thus restricted in their access patterns, i.e. loop carried dependencies are not supported) and loops that are always executed sequentially. The attributes \verb|domName| and \verb|domSize| specify the relevant domain iterators and the domain size relevant to data objects accessed within the parallel region (this relevancy will later be discussed in more detail in Section \ref{sec:hf-layout}). The attributes \verb|startAt| and \verb|endAt| explicitly state the region boundaries, which can be a subset of the domain size (however, if omitted, the domain size is also assumed as the region boundary).

\medskip

For CPU targets, Hybrid Fortran generates an OpenMP code version very similar to the reference code shown in Listing \ref{listing:lateral_damping}\footnote{Privatization is the main difference: Hybrid Fortran generated OpenMP code uses ``firstprivate'' as the default policy with an explicit ``shared'' clause for all arrays used in the kernel.}, with multi-core parallelization applied to the outermost loop. For GPU targets it defaults to CUDA Fortran kernels (thus generating all the necessary host- and device code boilerplate and data copy operations) with an option to use OpenACC kernels with CUDA compatible data structures (device pointers)\footnote{OpenACC is mainly used for reduction support - Hybrid Fortran does not automatically generate reduction kernels, however it supports the ``reduce'' clause, which is forwarded to the generated OpenMP or OpenACC kernels.}. The attribute \verb|template| specifies a macro suffix used for the generated block size parameters - this allows a central configuration for the block sizes used in an application, rather than leaking this architecture-specific optimization to the user code in each kernel. If omitted, configurable default block sizes are used.

\medskip

\begin{figure}[htpb]
  \centering
  \includegraphics[width=1.0\linewidth]{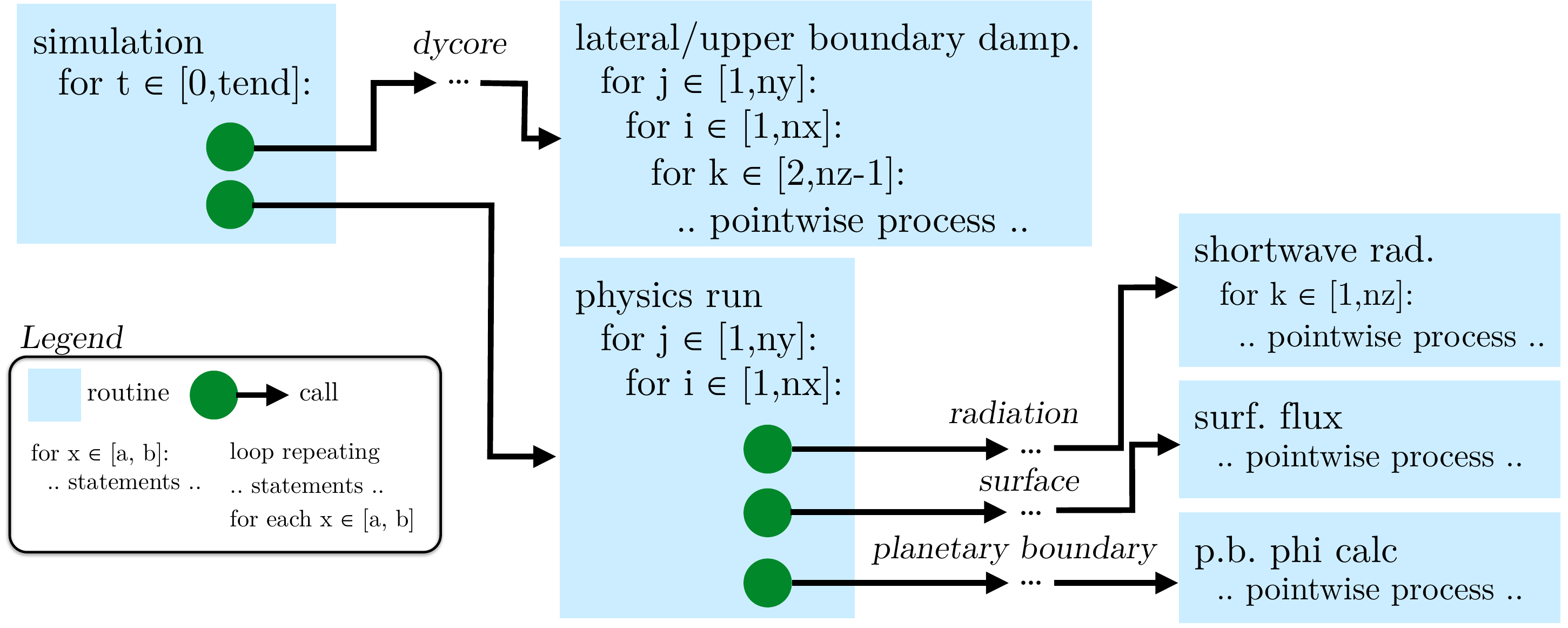}
  \caption{Simplified code structure of ASUCA.}
  \label{figure:asuca_structure}
\end{figure}

The main advantage of this parallelization DSL is the following: replacing parallelizeable loops with the \verb|@parallelRegion| construct allows the user to specify multiple granularities in the same code. Consider ASUCA's code structure, shown in simplified form in Figure \ref{figure:asuca_structure}. It shows two selected kernels and their embedding in the call graph - the lateral and upper boundary damping already discussed in this section, as well as the physics kernel. Many physical processes are called within this single kernel (of which three sample processes are depicted here). This code therefore has a very coarse granularity, which is problematic on GPU as discussed in Section \ref{sec:granularity}.

\medskip

\begin{figure}[htpb]
  \centering
  \includegraphics[width=1.0\linewidth]{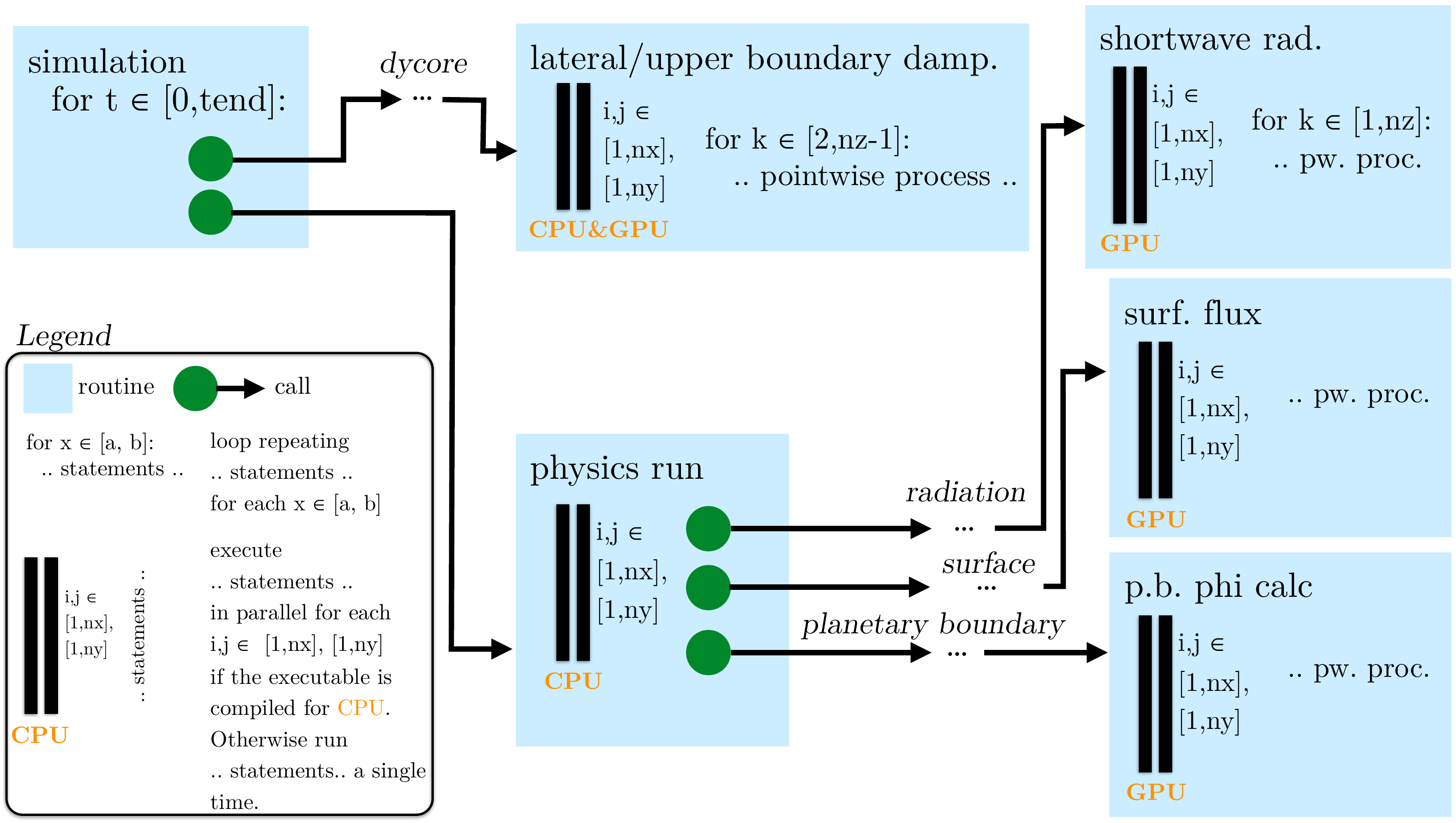}
  \caption{Simplified code structure of ASUCA using Hybrid Fortran.}
  \label{figure:asuca_structure_hf}
\end{figure}

With Hybrid Fortran we can solve this problem as follows: An additional \verb|appliesTo| attribute in the \verb|@parallelRegion| statement allows the user to selectively apply parallel regions to either CPU or GPU. Applying the parallelization at different granularities therefore becomes possible\footnote{thus obviating the need for code duplication and/or deep inlining of call trees}, by enabling user-steered kernel fission. Figure \ref{figure:asuca_structure_hf} shows the resulting code structure, with the physics run being split into many kernels for GPU while remaining a single coarse grained kernel for CPU. The later Listing \ref{listing:surface_flux} gives an example of how such a kernel fission works in practice.

\subsection{Compile-time Defined Memory Layout and Device Data Region} \label{sec:hf-layout}

As discussed in Section \ref{sec:layout}, it is necessary to consider two major aspects for implementing the memory layout: storage order on one hand and the compile-time defined granularity requiring a varying dimensionalities of data objects on the other. Part of Hybrid Fortran is an additional language extension, the \verb|@domainDependant| construct\footnote{Please note: While this paper follows American English, the Hybrid Fortran language extension has originally been developed following British English, which becomes apparent in the spelling of ``domainDependant'' (\url{https://en.oxforddictionaries.com/definition/dependant}). We consider support for the American English spelling of this directive as part of our future efforts.}, as a declarative way for the user to specify additional required information concerning data objects. This concerns memory layout as well as device memory operations, which will be discussed in this section.

\subsubsection{Storage Order}

Revisiting the code sample from Section \ref{sec:hf-parallelization}, the following Listing shows the specification of the routine implementing the discussed lateral and upper damping kernel:

\begin{lstlisting}[name=lateral_damping_spec, label=listing:lateral_damping_spec, caption={Routine implementing the lateral and upper damping kernel with Hybrid Fortran.}]
subroutine lateral_and_upper_damping()
  use ref, only : dens_ref_f
  use svar, only : dens_ptb_damp
  ! ... further imports omitted  
  implicit none
  
  @domainDependant{ &
    & attribute(autoDom, present), &
    & accPP(AT_TIGHT_STENCIL), domPP(DOM_TIGHT_STENCIL) &
  & }
  dens_ref_f, dens_ptb_damp
  @end domainDependant

  @domainDependant{ &
    & attribute(autoDom, present), &
    & accPP(AT4_TIGHT_STENCIL), domPP(DOM4_TIGHT_STENCIL) &
  & }
  dens_ptb_bnd
  @end domainDependant
  
  ! ... initialisation of tratio_bnd and mtratio_bnd omitted
  ! ... kernel omitted (already shown in listing 1.2)
end subroutine
\end{lstlisting}

This shows the specification of the local module data object \verb|dens_ptb_bnd| (density perturbation in the boundary layer) as well as the external module data objects \verb|dens_ref_f| (reference density) and \verb|dens_ptb_damp| (density perturbation in ASUCA grid).

\medskip

The \verb|autoDom| attribute is used to delegate the dimensions setup to the data object specification parser (which gathers this information in a separate pass from the source modules, here \verb|ref| and \verb|svar|), rather than having the user specify the dimensions explicitly again in the \verb|@domainDependant| construct. The attributes \verb|accPP| and \verb|domPP| are employed to specify the macro names used to implement the dimension ordering for accesses and specification parts, respectively. These macros wrap all dimension lists in access expressions and specifications of respective data objects in the generated code. When \verb|accPP| and \verb|domPP| attributes are omitted, default macro names are used (for a code example after this conversion please refer to Listing \ref{listing:surface_flux_openACC}). In case of Listing \ref{listing:lateral_damping_spec} we use explicit macro names for the dynamical core since the default macros are already used with different assumptions for the physical processes (see the paragraph on ``Dimensionality Changes'' below).

\subsubsection{Device Data Region}

Similar to OpenACC, in Hybrid Fortran we implement data regions by adding state attributes to data objects. The \verb|present| attribute, shown in Listing \ref{listing:lateral_damping_spec}, indicates that the respective objects are located on the device in case of GPU compilation. Analogous \verb|transferHere| attributes are used in the main simulation routine in order to instruct Hybrid Fortran to implement the memory copy operations to- and from the device, once at the beginning and end of the simulation. For dummy variables with specified \verb|intent|, Hybrid Fortran will use the Fortran intent information to determine the correct copy operation\footnote{Simple examples of this feature can be found in \url{https://github.com/muellermichel/Hybrid-Fortran/blob/v1.00rc10/examples/demo/source/example.h90}.}, which minimizes the potential for bugs in comparison to OpenACC's explicit \verb|copyIn|, \verb|copyOut| and \verb|copy| clauses. Halo region updates, required for every timestep, are implemented explicitly in code sections guarded from CPU compilation.

\subsubsection{Dimensionality Changes}

Due to the compile-time defined parallelization granularity, discussed in Section \ref{sec:hf-parallelization}, it is necessary to modify the dimensionality of data objects in certain cases in the source generation. This requires hints from the framework user. Consider the following surface flux code snippet:

\begin{lstlisting}[name=surface_flux, label=listing:surface_flux, caption={Surface flux code snippet.}]
lt = tile_land
if (tlcvr(lt) > 0.0_r_size) then
  call sf_slab_flx_land_run( &
    ! ... inputs and further tile variables omitted
    & taux_tile_ex(lt), tauy_tile_ex(lt) &
    & )

  u_f(lt) = sqrt(sqrt(taux_tile_ex(lt) ** 2 + tauy_tile_ex(lt) ** 2))
else
  taux_tile_ex(lt) = 0.0_r_size
  tauy_tile_ex(lt) = 0.0_r_size
  ! ... further tile variables omitted
end if
! ... sea tiles code and variable summing omitted
\end{lstlisting}

Since this process is defined inside the call graph of the physics kernel, as shown in Figure \ref{figure:asuca_structure}, the relevant 2D- and 3D grid point values are already sliced and passed in as scalars or 1D-arrays, that is, data parallelism is not exposed at this level. Hybrid Fortran allows implementing this as a fine-grained kernel (as outlined in Figure \ref{figure:asuca_structure_hf}) without modifying the computational user code, as demonstrated in the following snippet:

\begin{lstlisting}[name=surface_flux_hf, label=listing:surface_flux_hf, caption={Surface flux code snippet with Hybrid Fortran.}]
@domainDependant{domName(i,j), domSize(nx,ny), attribute(autoDom, present)}
tlcvr, taux_tile_ex, tauy_tile_ex, u_f
@end domainDependant

@parallelRegion{appliesTo(GPU), domName(i,j), domSize(nx,ny)}
lt = tile_land
if (tlcvr(lt) > 0.0_r_size) then
  call sf_slab_flx_land_run( &
    ! ... inputs and further tile variables omitted
    & taux_tile_ex(lt), tauy_tile_ex(lt) &
    & )

  u_f(lt) = sqrt(sqrt(taux_tile_ex(lt) ** 2 + tauy_tile_ex(lt) ** 2))
else
  taux_tile_ex(lt) = 0.0_r_size
  tauy_tile_ex(lt) = 0.0_r_size
  ! ... further tile variables omitted
end if
! ... sea tiles code and variable summing omitted
@end parallelRegion
\end{lstlisting}

Using our parallelization DSL to provide additional dimensionality information, Hybrid Fortran is able to rewrite this code into a 2D kernel. Dimensions missing from the user code are inserted at the beginning of the dimension lists in access expressions and data object specifications. As an example, the expression \verb|u_f(lt)| is converted to \verb|u_f(AT(i,j,lt))|, employing the default ordering macro already mentioned in the paragraph ``Storage Order''. Dimensions are extended whenever there is a match found for \verb|domName| or \verb|domSize| information between data objects and parallel regions within the same routine \textit{or} in routines called within the call graph of the same routine. It is therefore necessary for Hybrid Fortran to gather global information about the application before implementing each routine.

\subsection{Transformed Code} \label{sec:hf-transformed}

Revisiting Listing \ref{listing:surface_flux_hf}, the following code is generated when applying Hybrid Fortran with the OpenACC backend:

\begin{lstlisting}[name=surface_flux_hf, label=listing:surface_flux_openACC, caption={Surface flux code snippet after conversion with OpenACC backend.}]
!$acc kernels deviceptr(taux_tile_ex) deviceptr(tauy_tile_ex)  &
!$acc& deviceptr(tlcvr) deviceptr(u_f)
!$acc loop independent vector(CUDA_BLOCKSIZE_Y)
outerParallelLoop0: do j=1,ny
!$acc loop independent vector(CUDA_BLOCKSIZE_X)
	do i=1,nx
	    ! *** loop body *** :
		lt = tile_land
		if (tlcvr( AT(i,j,lt) )> 0.0_r_size) then
            call sf_slab_flx_land_run(&
                ! ... inputs and further tile variables omitted
                & taux_tile_ex( AT(i,j,lt) ), tauy_tile_ex( AT(i,j,lt) ) &
                & )
			u_f( AT(i,j,lt) )= sqrt(sqrt(taux_tile_ex( AT(i,j,lt) )** 2 + &
			    & tauy_tile_ex( AT(i,j,lt) )** 2))
		else
			taux_tile_ex( AT(i,j,lt) )= 0.0_r_size
			tauy_tile_ex( AT(i,j,lt) )= 0.0_r_size
			! ... further tile variables omitted
		end if
		! ... sea tiles code and variable summing omitted
	end do
end do outerParallelLoop0
!$acc end kernels
\end{lstlisting}

Device data is interoperable with the CUDA Fortran backend, thus device pointers are used instead of passing the management to OpenACC. OpenACC directives together with this data type can thus be directly used in the user code as well, i.e. it remains interoperable with device code generated by Hybrid Fortran.

\medskip

As noted in Section \ref{sec:hf-layout}, storage ordering macros (here \verb|AT()|) are applied to all array access statements. For the thread block setup, the configurable default sizes \verb|CUDA_BLOCKSIZE_X/Y| are used since no template is specified for the parallel region at hand. Parallel region loops (here for indices \verb|i| and |verb|j| are set up explicitly to parallelize. Other loops, such as the loop over \verb|k|, use a \verb|!$acc loop seq| directive to explicitly avoid parallization and give the framework user full expressiveness over the desired granularity.

\medskip

Applying CUDA Fortran backend to the same user code produces the following host code (here shown together with the routine header and footer):

\begin{lstlisting}[name=surface_flux_cuda_host, label=listing:surface_flux_cuda_host, caption={Surface flux host code snippet after conversion with CUDA Fortran backend.}]
subroutine hfd_sf_slab_flx_tile_run( &
	! ... inputs omitted
& )
	use cudafor
	type(dim3) :: cugrid, cublock
	integer(4) :: cugridSizeX, cugridSizeY, cugridSizeZ, &
	  & cuerror, cuErrorMemcopy
	! ... other imports and specifications omitted

	cuerror = cudaFuncSetCacheConfig( &
		& hfk0_sf_slab_flx_tile_run, cudaFuncCachePreferL1)
	cuerror = cudaGetLastError()
	if(cuerror .NE. cudaSuccess) then
		! error logging omitted
		stop 1
	end if
	cugridSizeX = ceiling(real(nx) / real(CUDA_BLOCKSIZE_X))
	cugridSizeY = ceiling(real(ny) / real(CUDA_BLOCKSIZE_Y))
	cugridSizeZ = 1
	cugrid = dim3(cugridSizeX, cugridSizeY, cugridSizeZ)
	cublock = dim3(CUDA_BLOCKSIZE_X, CUDA_BLOCKSIZE_Y, 1)
	call hfk0_sf_slab_flx_tile_run <<< cugrid, cublock >>>( &
		! ... inputs and further tile variables omitted
		& nx, ny, tile_land, u_f, tlcvr & ! required data objects are
		& taux_tile_ex, tauy_tile_ex &    ! automatically passed to kernel
		& )
	cuerror = cudaThreadSynchronize()
	cuerror = cudaGetLastError()
	if(cuerror .NE. cudaSuccess) then
		! error logging omitted
		stop 1
	end if
end subroutine
\end{lstlisting}

The prefix \verb|hfd_| is added to host routines that use device data. Hybrid Fortran also duplicates the code for a pure host version of these routines (without a name change in order to remain interoperable with code that is not passed through Hybrid Fortran). In contexts where the data is not residing on the device, such as the setup part of an application, Hybrid Fortran automatically chooses the host version when generating the call statements at compile-time. Code residing within parallel regions is moved within a separeted kernel routine (using prefix \verb|hfki_| with \verb|i| representing the kernel number). In case of the surface flux sample shown here, the kernel routine is generated as follows:

\begin{lstlisting}[name=surface_flux_cuda_device, label=listing:surface_flux_cuda_device, caption={Surface flux device code snippet after conversion with CUDA Fortran backend.}]
attributes(global) subroutine hfk0_sf_slab_flx_tile_run(&
	! ... inputs and further tile variables omitted
	&, nx, ny, tile_land, u_f, tlcvr &
	&, taux_tile_ex, tauy_tile_ex &
	& )
	use cudafor
	use pp_vardef ! defines r_size
	implicit none
	real(r_size), device :: u_f(DOM(nx,ny,ntlm))
	real(r_size), device :: tlcvr(DOM(nx,ny,ntlm))
	real(r_size), device :: taux_tile_ex(DOM(nx,ny,ntlm))
	real(r_size), device :: tauy_tile_ex(DOM(nx,ny,ntlm))
	integer(4), value :: lt
	integer(4), value :: nx
	integer(4), value :: ny
	integer(4), value :: tile_land
	! ... other imports and specifications omitted

	i = (blockidx%x - 1) * blockDim%x + threadidx%x + 1 - 1
	j = (blockidx%y - 1) * blockDim%y + threadidx%y + 1 - 1
	if (i .GT. nx .OR. j .GT. ny) then
		return
	end if
	! *** loop body *** :
	lt = tile_land
	if (tlcvr( AT(i,j,lt) )> 0.0_r_size) then
		call hfd_sf_slab_flx_land_run( &
			! ... inputs and further tile variables omitted
			taux_tile_ex( AT(i,j,lt) ), tauy_tile_ex( AT(i,j,lt) ) &
			& )
		! ... rest of loop body already shown in listing 1.6
end subroutine
\end{lstlisting}

The specification part of these kernels is automatically generated, applying device state information and converting input scalars to pass-by-value\footnote{reduction kernels are thus not supported with this backend - we use the OpenACC backend selectively for this purpose, see also the discussion in the footnotes to Section \ref{sec:hf-parallelization}.}, among other transformations.

\medskip

It is notable that CUDA Fortran requires a fairly large amount of boiler plate code for grid setup, iterator setup, host- and device code separation as well as memory- and error handling - Hybrid Fortran allows the user to pass on the responsibility for that to the framework. Compared with the code generated by OpenACC however (assembly-like CUDA C or NVVM intermediate representation), the Hybrid Fortran generated CUDA Fortran code remains easily readable to programmers experienced with CUDA. Experience shows that this is a productivity boost, especially in the debugging and manual performance optimization phase of a project.

\medskip

Regarding the OpenMP backend, since the surface flux example is parallelized at a much more  coarse-grained level for CPU, the generated CPU code for the sample at hand is a one-to-one copy of the user code shown earlier in Listing \ref{listing:surface_flux}. The parallelization is generated at a higher level in the call graph (replacing the parallel region construct with \verb|appliesTo(CPU)| clause) as follows:

\begin{lstlisting}[name=surface_flux_openmp, label=listing:surface_flux_openmp, caption={Surface flux device code snippet after conversion with CUDA Fortran backend.}]
!$OMP PARALLEL DO DEFAULT(firstprivate)  
!$OMP& SHARED( ... inputs and outputs omitted ... )
		outerParallelLoop0: do j=1,ny
		do i=1,nx
			call physics_main(i, j, &
			  ! ... inputs and outputs omitted
			  & )
		end do
	end do outerParallelLoop0
!$OMP END PARALLEL DO
\end{lstlisting}

\section{Code Transformation Method} \label{sec:hf-transformation}

\begin{figure}[htpb]
  \centering
  \includegraphics[width=1.0\linewidth]{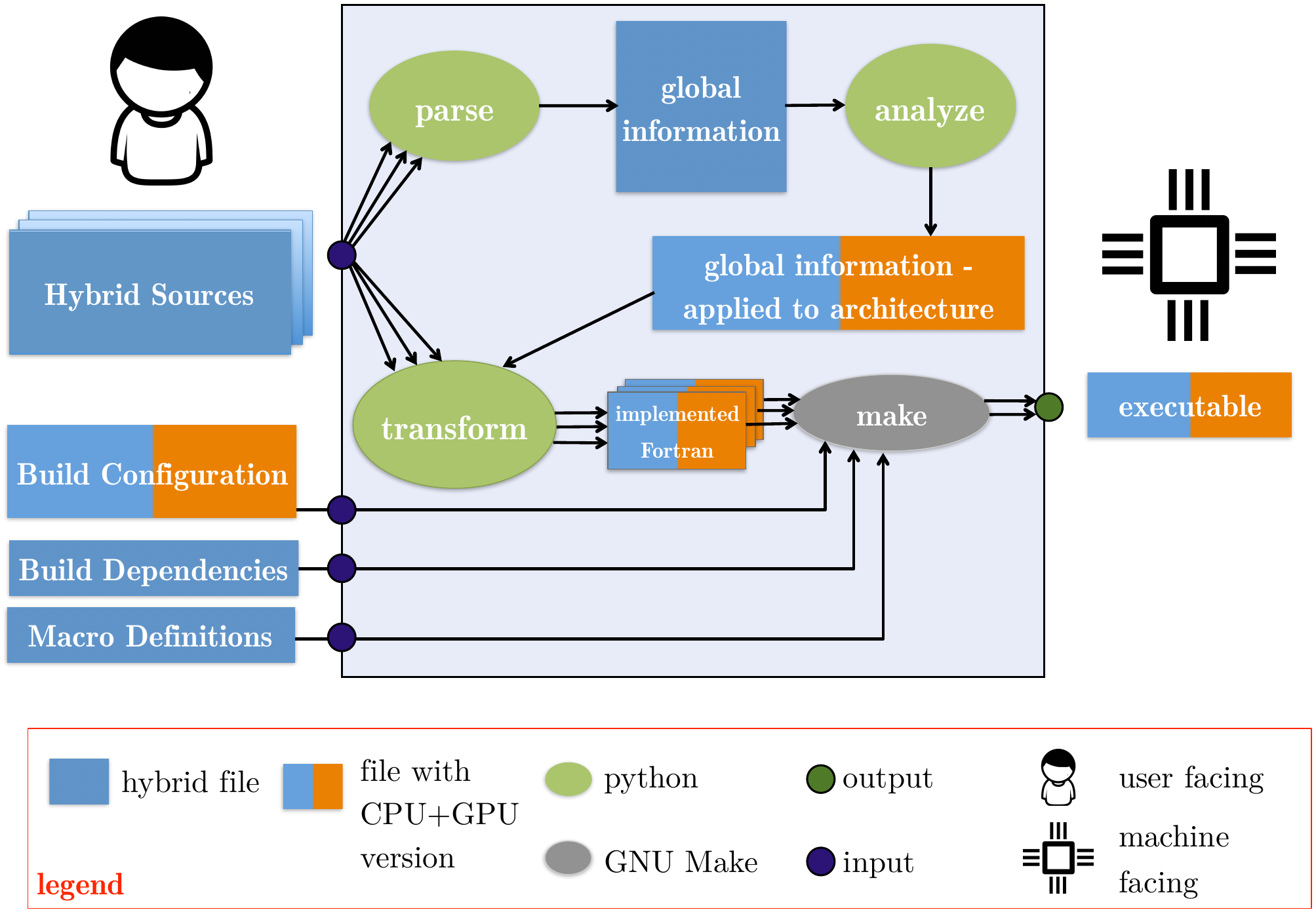}
  \caption{Hybrid Fortran software components and build workflow.}
  \label{figure:hf_schematics}
\end{figure}

In this section we discuss code transformation method involved in implementing Hybrid Fortran's characteristics described earlier. This process is applied transparently for the user, i.e. it is applied automatically by the means of a provided common Makefile\footnote{See also the ``Getting Started'' section in \url{https://github.com/muellermichel/Hybrid-Fortran/blob/v1.00rc10/doc/Documentation.pdf}.}. Figure \ref{figure:hf_schematics} gives an overview of the process and the components involved. We discuss this process in order of execution - each of the following enumerated items corresponds to one transformation phase:

\begin{enumerate}
 \item To simplify the parsing in subsequent phases, Fortran continuation lines are merged.
 \item Facilitating later phases, the application's call graph and parallel region directives are parsed globally (``parse'' phase in Figure \ref{figure:hf_schematics}).
 \item Using the \verb|appliesTo| information in kernels and the call graph, the position of each routine in relation to kernels is computed.
 Possible positions are ``has kernel(s) in its caller graph'', ``contains a kernel itself'' and ``is called inside a kernel'' (``analyze'' phase in Figure \ref{figure:hf_schematics}).
 \item In two passes, module data object specifications are parsed and then linked against all routines with imports of such objects, together with the locally defined objects.
 \item A global application model is generated, with model classes representing the modules, routines and code regions.
This model can be regarded as a target hardware independent intermediate representation.
 \item Each routine object is assigned an implementation class depending on the target architecture\footnote{Hybrid Fortran allows the user to switch between varying backend implementations per routine, such as OpenACC and CUDA Fortran - the user specified information as well as the defaults given by the build system call thus steers this implementation class.}. For each coding pattern, a separate class method of the implementation class is called by the model objects - e.g. CUDA parallelization boilerplate is generated.
Using the previously gathered global kernel positioning and data object dimension information, data objects are transformed according to the behavior discussed in Section \ref{sec:hf-layout}. Implementation class methods return strings that are concatenated by the model objects into source files (``transform'' phase in Figure \ref{figure:hf_schematics}).
 \item Code lines are split using Fortran line continuations in order to adhere to line limits imposed by Fortran compilers.
 \item Macros generated by Hybrid Fortran (to implement storage reordering and configurable block sizes) are processed by using the GNU compiler toolchain. Subsequently, a user specified compiler and linker is employed in order to create the CPU and GPU executables. A common makefile is provided with the framework, however the build dependency graph is user-provided in the format of makefile rules\footnote{alternatively, a dependency generator script can be configured as well} (``make'' phase in Figure \ref{figure:hf_schematics}).
\end{enumerate}

This process makes it possible to have a unified source input and create executables targeted for either multi-core CPU or many-core GPU.
\section{Productivity- and Performance Results} \label{sec:results}

In this section we discuss the productivity- and performance results that have been achieved when applying Hybrid Fortran to weather prediction models.

\medskip

At the time of this writing an additional paper has been submitted for review, in which the following results are discussed: Hybrid Fortran has been applied to both dynamical core and physical processes of ASUCA. This implementation (here referred to as ``Hybrid ASUCA'') consists of 338 kernels and covers almost all modules required for an operative weather prediction, with convection being the main exception due to development time constraints. Communication between host and device has been eliminated with the exception of setup, output and halo exchange. On a 301 x 301 x 58 grid with real weather data, a 4.9x speedup has been achieved when comparing the Hybrid Fortran port on four Tesla K20x versus the JMA provided reference implementation on four 6-core Westmere Xeon X5670 (TSUBAME 2.5). The same setup executes at a speedup of 3x when comparing a single Tesla P100 versus a single 18-core Broadwell Xeon E5-2695 v4 (Reedbush-H). On a full-scale production run with 1581 x 1301 x 58 grid size and 2km resolution, 24 Tesla P100 GPUs are shown to replace more than 50 18-core Broadwell Xeon sockets \cite{mueller2017unpub}.

\medskip

In order to further examine the performance of Hybrid Fortran kernels and compare them to OpenACC and OpenMP user codes, a performance model has been constructed for a reduced weather application. The Hybrid Fortran implementation with unified code performs on par or better than OpenACC- and OpenMP- implementations separately optimized for GPU and CPU, respectively. In this application, whose runtime is dominated by the memory bandwidth used for a seven point stencil diffusion kernel, the Hybrid Fortran GPU version performs 25\% better than the no-cache-model on Tesla K20x. It achieves 36\% of the theoretical limit given by a perfect-cache-model. The CPU version performs 56\% better than the no-cache-model and achieves 67\% of the theoretical limit on Westmere Xeon X5670 \cite{mueller2017unpub}.

\medskip

\begin{figure}[htpb]
  \centering
  \includegraphics[width=0.85\linewidth]{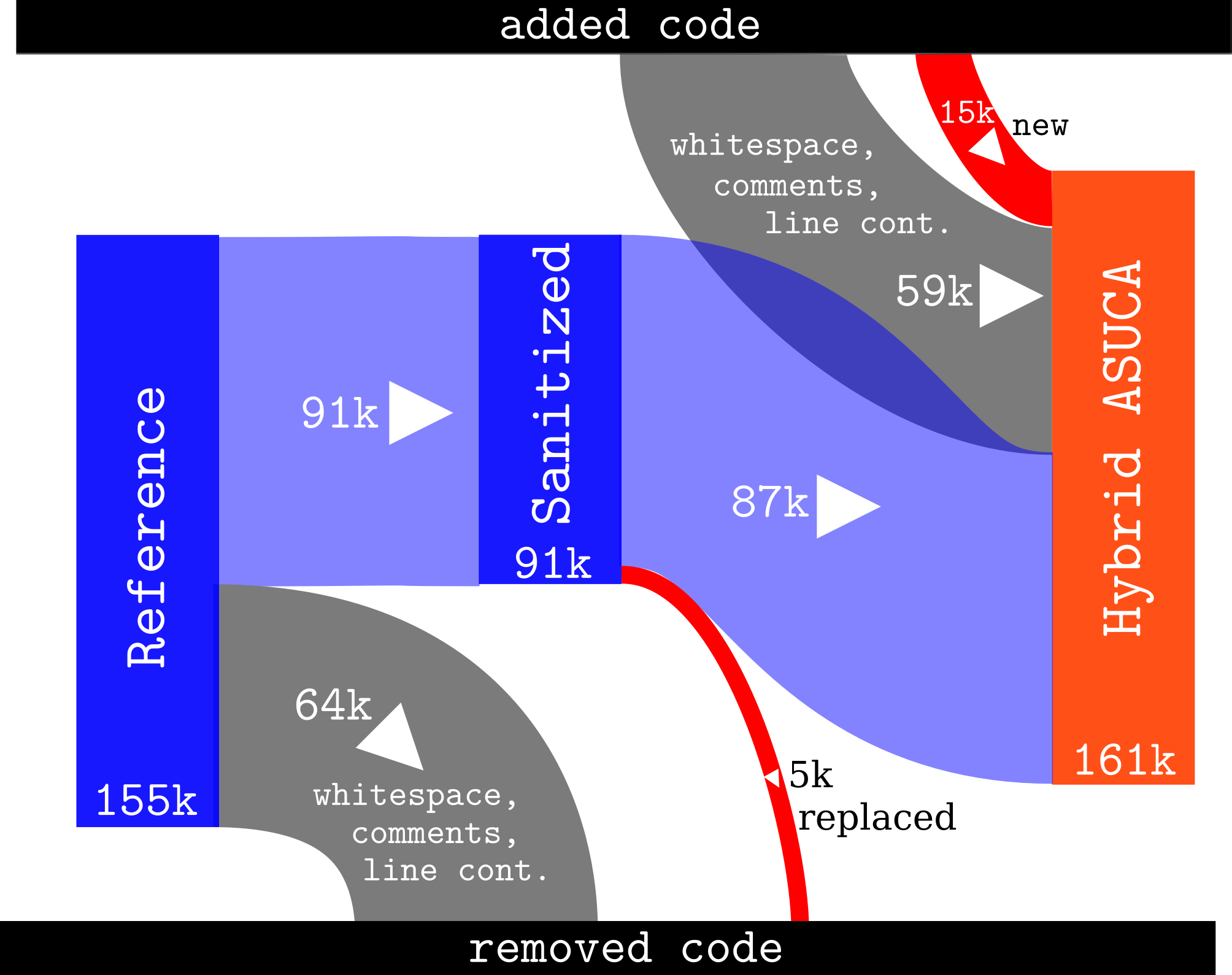}
  \caption{Flow of code lines from reference implementation to Hybrid ASUCA by number of lines of code.}
  \label{figure:code-flow}
\end{figure}

To examine the productivity of our solution we have analyzed the code and compare it against the reference implementation\footnote{Since the input to this analysis is the closed source ASUCA codebase, full reproducibility cannot be provided in this context. However the intermediate data, the method employed to gather this data as well as a sample input is provided and documented in \url{https://github.com/muellermichel/hybrid-asuca-productivity-evidence/blob/master/asuca_productivity.xlsx}.}. The high-level results of this analysis is shown in Figure \ref{figure:code-flow}. In order to gain GPU support in addition to the already existing multi-core and multi-node parallelization, the code has grown by less than 4\% in total, from 155k lines of code to 161k. Sanitizing the two code versions (removing white space, comments and merging continued lines), the code has grown by 12\%, from 91k to 102k lines of code. 95\% of the sanitized reference code is used as-is in the new implementation, while 5\% or approximately 5k lines of code is replaced with approximately 15k new code lines.

\medskip

\begin{figure}[htpb]
  \centering
  \includegraphics[width=0.8\linewidth]{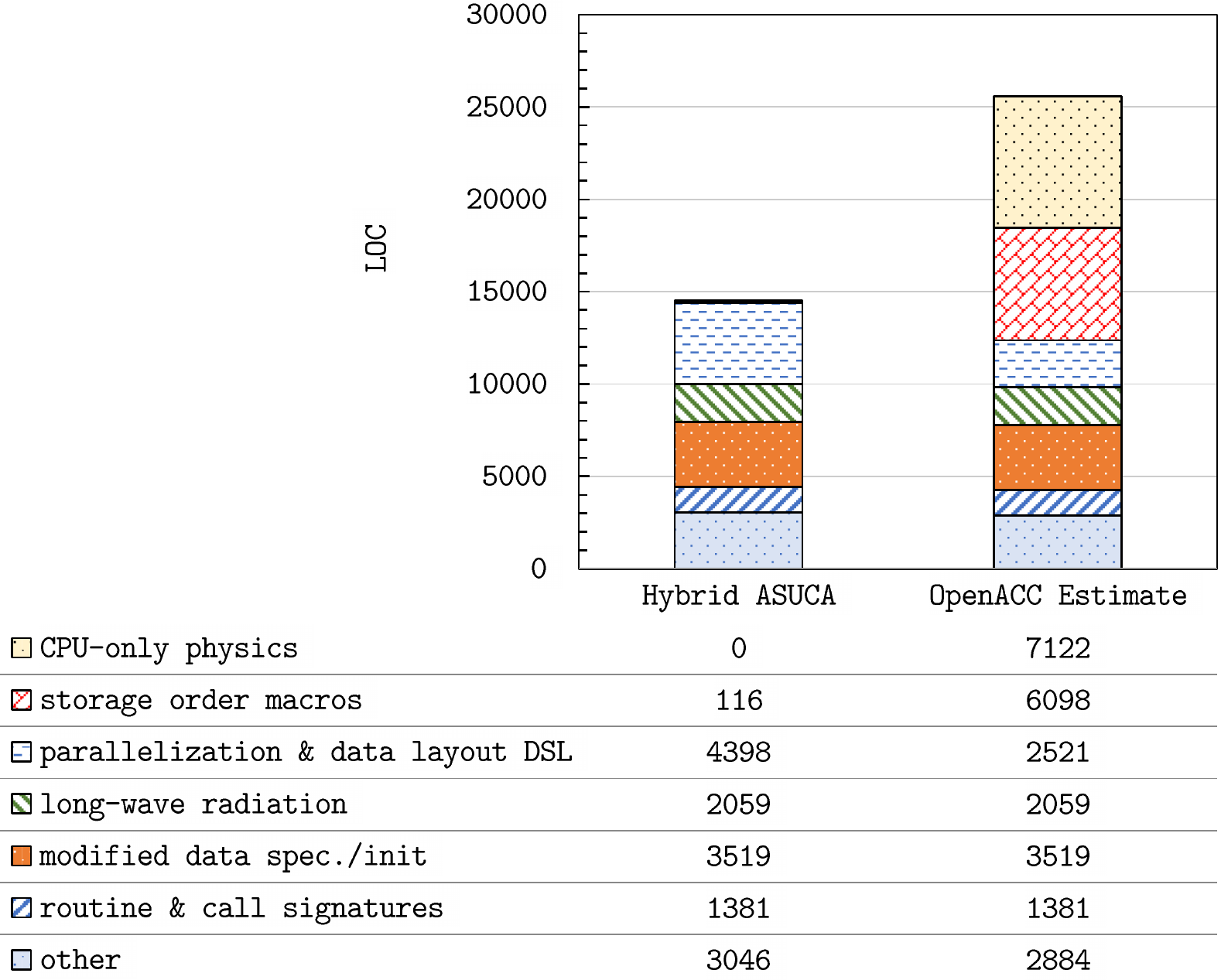}
  \caption{New code required for Hybrid ASUCA vs. estimate of equivalent OpenACC implementation (LOC stands for ``lines of code'').}
  \label{figure:changed-code}
\end{figure}

Code changes and additions have the largest impact in terms of productivity. We have analyzed the additional 15k lines of code in more detail. Figure \ref{figure:changed-code} shows a breakdown of these changes and compares them to an estimate of what would be required with an OpenACC-based implementation. The following methodology has been used for this analysis:

\begin{enumerate}
\item for the parallelization- and data layout DSL line count we have used information parsed for Hybrid ASUCA, as well as the OpenACC backend available in Hybrid Fortran, to acquire an accurate count for the directives required for an OpenACC-based implementation,
\item since OpenACC does not offer a granularity abstraction, we have used the Hybrid ASUCA's parsed global application information to arrive at a set of routines that require a kernel positioning change (see the discussion in Section \ref{sec:hf-transformation}) - the resulting code lines require duplication for the CPU in an equivalent OpenACC implementation (shown as ``CPU-only physics'' in figure \ref{figure:changed-code}),
\item we have used the OpenACC backend in Hybrid Fortran to count the lines of code where storage order macros are introduced in order to achieve a compile-time defined data layout.
\end{enumerate}

``Parallelization \& data layout DSL'' refers to the number of code lines for \verb|@parallelRegion| and \verb|@domainDependant| directives in case of Hybrid Fortran, and OpenACC \verb|!$acc| directivies in case of OpenACC. Hybrid Fortran replaces the requirement for code changes to implement a varying data layout (``storage order macros'', 6098 lines of code, ``LOC'') as well as a code duplication for multiple parallelizations with varying granularities (``CPU-only physics'', 7122 LOC) with a higher number of DSL code lines compared to OpenACC (4398 vs. 2521 LOC). ``Modified data specifications / initializations'' (3519 LOC) as well as ``routine \& call signature'' (1381 LOC) refers to changes applied to the setup of data and call parameter lists, respectively. These changes are necessary due to device code limitations and optimizations and are largely required for both the Hybrid Fortran version as well as a potential OpenACC solution, so we use the result from Hybrid ASUCA as an estimate for what would be required with OpenACC. Finally, one physics module concerning long-wave radiation (2059 LOC), has been replaced with a version that uses less local memory per thread (a factor of 10 improvement in that regard) to make it more GPU-friendly. We again estimate that an OpenACC version would have approximately the same code size.

\medskip

This result shows that an equivalent OpenACC implementation of ASUCA can be estimated to require approximately 11k LOC in additional changes compared to the Hybrid Fortran-based implementation. When comparing to the sanitized reference codebase, an OpenACC user code would require approximately 28\% of code lines to be changed or added, while Hybrid Fortran currently requires 16\%.

\medskip

In addition to the results regarding ASUCA presented here, a library of small applications and kernel samples has been created to demonstrate Hybrid Fortran general applicability to data parallel code\footnote{Please refer to \url{https://github.com/muellermichel/Hybrid-Fortran/blob/v1.00rc10/examples/Overview.md} for an overview of the available samples and their results.}.
\section{Conclusion and Future Work}\label{sec:conclusions}

With this work we have shown that it is possible for large structured grid Fortran applications to

\begin{enumerate}
\item achieve a GPU implementation without rewriting major parts of the computational code,
\item abstract the memory layout and
\item allow for multiple parallelization granularities.
\end{enumerate}

With our proposed method, a regional scale weather prediction model of significant importance to Japan's national weather service has been ported to GPU, showing a speedup of up to 4.9x on single GPU compared to a single Xeon socket. When scaling up to 24 Tesla P100, less than half the number of GPUs is required compared to contemporary Xeon CPU sockets to achieve the same result. Approximately 95\% of the existing codebase has been reused for this implementation and our implementation has grown by less than 4\% in total, even though it is now supporting GPU as well as CPU. Through a library of results, the general applicability of our framework to data parallel Fortran code has been shown.

\medskip

Considering that much of the changes still required in the user code are ``mechanical'' in nature, we expect additional productivity gains to be possible from further automation. We strive to achieve a solution where a Hybrid Fortran-based transformation can be applied to large structured grid applications wholesale with minimal input required by the user.

\subsubsection*{Acknowledgments}
This work has been supported by the Japan Science and Technology Agency (JST) Core Research of Evolutional Science and Technology (CREST) research program ``Highly Productive, High Performance Application Frameworks for Post Peta-scale Computing'', by KAKENHI Grant-in-Aid for Scientific Research (S) 26220002 from the Ministry of Education, Culture, Sports, Science and Technology (MEXT) of Japan, by ``Joint Usage/Research Center'' for Interdisciplinary Large-scale Information Infrastructures (JHPCN)" and ``High Performance Computing Infrastructure (HPCI)'' as well as by the ``Advanced Computation and I/O Methods for Earth-System Simulations'' (AIMES) project running under the German-Japanese priority program ``Software for Exascale Computing'' (SPPEXA). The authors thank the Japan Meteorological Agency for their extensive support, Tokyo University and the Global Scientific Information and Computing Center at Tokyo Institute of Technology for the use of their supercomputers Reedbush-H and TSUBAME 2.5.

\bibliographystyle{splncs03}
\bibliography{content/references}


\end{document}